# Towards Recommender Systems LLMs Playground (RecSysLLMsP): Exploring Polarization and Engagement in Simulated Social Networks*

Utilizing Simulation Software to Evaluate the Impact of Recommender Systems on Social Networks

Ljubiša Bojić[1,2,*,†], Zorica Dodevska[1,†], Yashar Deldjoo[3,†] and Nenad Pantelić[4,†]

[1]*Institute for Artificial Intelligence Research and Development of Serbia, 1 Fruškogorska, Novi Sad, 21000, Serbia*
[2]*University of Belgrade, Institute for Philosophy and Social Theory, 45 Kraljice Natalije, Belgrade, 11000, Serbia*
[3]*Polytechnic University of Bari, 4 Via Orabona, Bari, 70125, Italy*
[4]*University of Kragujevac, Faculty of Engineering, 6 Sestre Janjić, Kragujevac, 34000, Serbia*

**Abstract**

Given the exponential advancement in AI technologies and the potential escalation of harmful effects from recommendation systems, it is crucial to simulate and evaluate these effects early on. Doing so can help prevent possible damage to both societies and technology companies. This paper introduces the Recommender Systems LLMs Playground (RecSysLLMsP), a novel simulation framework leveraging Large Language Models (LLMs) to explore the impacts of different content recommendation setups on user engagement and polarization in social networks. By creating diverse AI agents (AgentPrompts) with descriptive, static, and dynamic attributes, we assess their autonomous behaviour across three scenarios: Plurality, Balanced, and Similarity. Our findings reveal that the Similarity Scenario, which aligns content with user preferences, maximizes engagement while potentially fostering echo chambers. Conversely, the Plurality Scenario promotes diverse interactions but produces mixed engagement results. Our study emphasizes the need for a careful balance in recommender system designs to enhance user satisfaction while mitigating societal polarization. It underscores the unique value and challenges of incorporating LLMs into simulation environments. The benefits of RecSysLLMsP lie in its potential to calculate polarization effects, which is crucial for assessing societal impacts and determining user engagement levels with diverse recommender system setups. This advantage is essential for developing and maintaining a successful business model for social media companies. However, the study's limitations revolve around accurately emulating reality. Future efforts should validate the similarity in behaviour between real humans and AgentPrompts and establish metrics for measuring polarization scores. Additionally, the model's realism can be enhanced by incorporating various types of content into the simulation and expanding the range of AgentPrompts in actual experiments. This research direction paves the way toward developing and validating more comprehensive simulation software. An additional benefit of this software is the potential to create novel recommender systems based on LLMs and produce synthetic data for further training these models.

## 1. Introduction

The advent of the digital age and the emergence of social media platforms have revolutionized how we socialize, communicate, and, more importantly, consume information. However, along with the many conveniences of this digital revolution, several concerns necessitate exhaustive exploration. One such significant concern revolves around the alleged implications of social media algorithms and their effects on social polarization, aiding populist leaders and propagating false information [1, 2].

Our societies increasingly rely on social media for news and information, changing how we form opinions and make decisions [3]. Much of this influence is attributed to recommender systems – the algorithms that decide and filter what we see based on user preferences and past digital behaviour [4, 5, 6]. Given the predominantly commercial intent of these systems, the focus is mainly on engaging and retaining user attention [7, 8]. Therefore, it gives rise to 'filter bubbles' – where users are repetitively exposed to similar content, reinforcing initially held beliefs and possibly sourcing in echo chambers and polarization [9]. One kind of recommender system polarization is group unfairness based on binary-defined attributes [10]. Societal risks regarding biases, 'filter bubbles' and echo chambers associated with recommender systems are already recognized topics of concern, especially when they come with generative models [11, 12].

Multiple studies have revealed substantial evidence supporting these allegations, with populist leaders accused of exploiting these algorithms to manipulate social networks for political gains [13, 14]. An imbalanced exposure to information further perpetuates social polarization, giving rise to distinct ideological clusters [15, 16, 17, 18].

From the perspective of social media companies, as personalized content and recommender systems are lucrative means of user engagement and retention, it could be helpful to simulate different effects of recommender systems setups on user engagement before they take place in real life, as this could save lots of assets and sustain business models [19, 20]. Social media companies also need to analyze and revise their algorithms to prevent the spread of false information and mitigate social polarization. However, such revisions need diligent testing in artificial environments before real-world applications, failing which unanticipated consequences may arise [21].

Given these substantial social and economic implications,

a practical solution that allows a systematic, comprehensive, yet secure way to test the impact of various recommender system setups is desirable. Therefore, our study proposes the development of a novel simulation software, Recommender Systems Playground (RecSysLLMsP). It aims to simulate a social network environment in an artificial setting using Large Language Models (LLMs) to create numerous unique artificial intelligence (AI) agents (i.e., AgentPrompts).

Incorporating LLMs within RecSysLLMsP offers significant advantages over traditional models. For instance, LLMs can generate more diverse and realistic modelling of user behaviours due to their ability to simulate understanding and process natural language at a high level [22]. Unlike rule-based systems or simpler machine learning models, LLM-based agents can exhibit diverse and unique interactions, making the simulation closer to real-world scenarios.

Despite their advanced capabilities, LLMs are susceptible to biases inherent in their training data [23, 24, 25, 26].

This property can lead to skewed interactions and outcomes within the simulation. Mitigation strategies include using diverse training datasets and implementing agent-based approaches where biases can be monitored and adjusted. Initiatives such as ensuring demographic diversity and incorporating fairness constraints could be used to improve the robustness of these models. Additionally, LLM platforms already invest significant efforts in LLM training processes to avoid biases in the first place.

Given these substantial social and economic implications, our study addresses the following research questions:

- How do variations in recommender system setups affect community engagement and polarization within a simulated social network?
- What are the broader societal implications of these changes?

This paper's primary contributions include the development of novel simulation software, Recommender Systems LLMs Playground (RecSysLLMsP), the use of LLMs to create diverse AI agents for simulation, as proposed in a theoretical framework [27], and the exploration of three distinct recommender system scenarios to measure their impact on social polarization and engagement. To answer the research questions, we propose a mathematical model to define AgentPrompts, capture the relationship among recommender systems and AI-agent interactions, and assess the potential effects on polarization and engagement.

## 2. Main Aspects of the RecSysLLMsP

The proposed methodology comprises four components: AgentPrompts, Recommender System Setup, PrimaryContent, and Interaction Rules.

AgentPrompts refer to unique AI agents who simulate real users in the social network. These agents are encoded with a set of assigned attributes and descriptive characteristics. These attributes are represented at 7-point Likert scales to simulate real-life users with different personality traits and dynamic preferences. A sample selection of agent prompts uniquely defines their personality and preference characteristics, which allows the model to generate diversity in agent behaviour and interaction patterns.

Recommender System Setup operates in three adjustable setups:

- Displaying a wide variety of content diverging from a user's interest and attitudes;
- Providing balanced content based on the user's interest and attitudes; and
- Presenting content aligning entirely with the user's interests and attitudes.

This study explores these setups' effects on community polarization and user engagement.

PrimaryContent denotes a consistent subject matter, such as the COVID-19 pandemic or the Middle Eastern conflict, that persists throughout a simulation. Users select these topics before launching the simulation. They serve as key factors in assessing the variation in community responses based on the configuration of the recommender system. However, additional content (posts) may originate from three main sources:

- friend connections (to the AgentPrompt),
- trending posts,
- imposed topics (or promotional content).

The AgentPrompt's interests drive the creation of varying profiles designed to mimic brands or lifestyle platforms capable of generating content relevant to the AgentPrompt yet not related to the PrimaryContent. Each AgentPrompt receives a blend of posts from these three sources daily, with the number of posts consumed individually determined for each AgentPrompt daily.

Interaction Rules' guidelines will be predefined, replicating the natural dynamics associated with social media engagement. AgentPrompts will engage with content following specific rule-based parameters that apply realistic interaction restrictions. An example of such rules may involve an AgentPrompt deciding whether to engage with each post from the daily mix.

If the AgentPrompt opts to interact, further decisions are made involving liking, commenting or resharing the post. Whether they read comments or not is also determined. If comments are read, the AgentPrompt additionally decides whether to comment and, if so, formulates the content of the remark. Finally, AgentPrompt generates an individual post that can be displayed to connections or appear in trending content for non-connections. Figure 1 illustrates the processes of the suggested RecSysLLMsP with an artificial AI agent (i.e., AgentPrompt) in the middle.

**Figure 1:** Processes of the suggested RecSysLLMsP

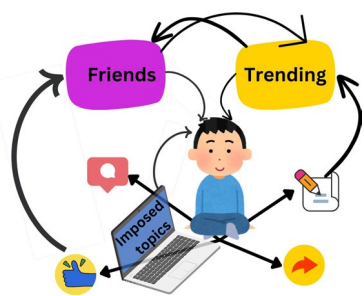

## 3. Definition of AgentPrompts

Each AgentPrompt has a friend list which can be expanded by sending friend requests when interacting with trending content or while reading comments on a post. The daily time each AgentPrompt spends on social media dictates the number of posts they can consume. The AgentPrompt's active social media duration moulds their daily browsing capacity. We define AgentPrompts over several dimensions described in Table 1.

**Table 1**
AgentPrompts' dimensions

| Dimension type | Dimension name | Abbrev. | Description | Scale range |
| --- | --- | --- | --- | --- |
| Descriptive | Nickname | *Nick* | A nickname or short name of the individual | *N/A* |
| Descriptive | Brief bio | *Bio* | A brief biography of the individual | *N/A* |
| Descriptive | List of interests | *Interests* | A list of the individual's interests | *N/A* |
| Static | Openness | $o$ | Personality trait measuring openness to experience | 1 (*very low*) to 7 (*very high*) |
| Static | Conscientiousness | $c$ | Personality trait measuring conscientiousness | 1 (*very low*) to 7 (*very high*) |
| Static | Extraversion | $e$ | Personality trait measuring extraversion | 1 (*very low*) to 7 (*very high*) |
| Static | Agreeableness | $a$ | Personality trait measuring agreeableness | 1 (*very low*) to 7 (*very high*) |
| Static | Neuroticism | $n$ | Personality trait measuring neuroticism | 1 (*very low*) to 7 (*very high*) |
| Static | Cognitive style | $cs$ | Measure of thinking style | 1 (*very analytical*) to 7 (*very emotional*) |
| Static | Open-mindedness | $om$ | Measure of openness of mind | 1 (*very closed-minded*) to 7 (*very open-minded*) |
| Dynamic | Political attitude | $pa$ | Measure of political orientation | 1 (*extremely liberal*) to 7 (*extremely conservative*) |
| Dynamic | Social connectivity | $sc$ | Measure of social connectivity | 1 (*very low connectivity*) to 7 (*very high connectivity*) |
| Dynamic | Emotive reaction | $er$ | Measure of level of emotional response | 1 (*very low emotional response*) to 7 (*very high emotional response*) |

Together, these attributes can provide a diversified pool of AgentPrompts for a comprehensive simulation. Each aspect can be adjusted and randomized to create unique AgentPrompts, simulating the diversity found in real-world social networks. We define AgentPrompt as a combination of descriptive dimensions (nickname, brief bio, and list of interests) and 10-dimensional vector dimensions ($P_i$, let's consider set size $i = 1, 2, ..., 10000$). Each vector dimension corresponds to one of the scaled characteristics described in Table 1, i.e.:

$$P_i = (o, c, e, a, n, cs, om, pa, sc, er) \quad (1)$$

This configuration of AgentPrompt allows us to have a mathematical representation of various AgentPrompts and better understand their interactions with each other and the content they interact with. It consists of 10 distinct measurable dimensions divided into the following types:

- The static type covers a range of personality traits such as openness ($o$), conscientiousness ($c$), extraversion ($e$), agreeableness ($a$), and neuroticism ($n$), rated on a scale from 1 (*very low*) to 7 (*very high*). It also includes a cognitive style ($cs$) measure, ranging from 1 for *very analytical thinking* to 7 for *very emotional thinking*, and an open-mindedness ($om$) measure, which spanned from 1 for *very closed-minded* to 7 for *very open-minded*.
- The dynamic type encapsulates the political attitude ($pa$) measure, ranging from 1 for *extremely liberal* to 7 for *extremely conservative*, along with the social connectivity ($sc$) and emotive reaction ($er$) measures, graded from 1 (*very low*) to 7 (*very high*). The dynamic part of an AgentPrompt evolves as the agent interacts with various types of content throughout the simulation. Changes in the values of dynamic dimensions represent this evolution. The dynamic nature of these attributes allows RecSysLLMsP to simulate realistic changes in user behaviour and preferences, capturing the complexity of social interactions and the impact of different recommender system setups on them.

## 4. Simplified Mathematical Model of a RecSysLLMsP Simulation

For the simulation, we will propose a mathematical model capturing the relationships between the recommender systems and AI Agent interactions and assess the potential effects on polarization and engagement. Given the possible scenarios and setups, we can illustrate the model with three generic situations for each recommender system setup specified above.

- Plurality Scenario: Recommender system setup focuses on exposing users to a broad array of content diverging from their pre-set interests and attitudes. In this scenario, we measure the effects of the recommender systems based on the shifts in AgentPrompts' preferences, engagement, polarization, and willingness to continue using the network.
- Balanced Scenario: In this setup, the recommender system introduces a balanced mix of content - few aligned with user interests and attitudes, and others diverging from it. The AgentPrompts' reactions and engagement with shared content will be observed, as well as any variations in their preferences and network usage willingness.
- Similarity Scenario: The recommender system presents content that reflects the user's pre-set interests and attitudes. This scenario aims to measure whether similarity in content reduces polarization, increases user engagement, or further amplifies user biases or polarizes them.

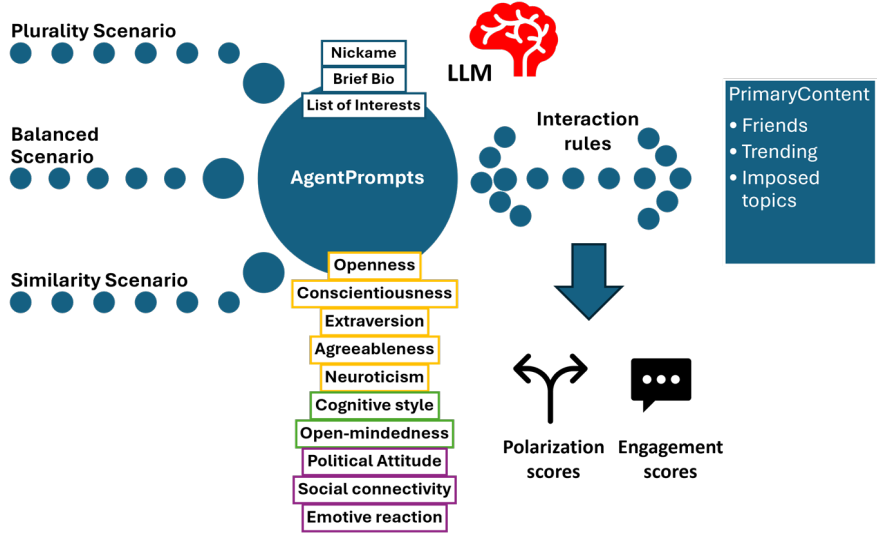

**Figure 2:** RecSysLLMsP's architecture

In each scenario (Figure 2), we analyze changes in the AgentPrompts' behaviour, preferences, and willingness to continue using the network to assess the impact of the selected recommender system. The outputs of this mathematical model should provide valuable insights into managing and setting up recommender systems for optimal societal impact and satisfaction.

A mathematical model to simulate this interaction will require us to define a few variables formally. Let $i$-th AgentPrompt ($P_i$) has a polarization score ($Ps_i$) and an engagement score ($Es_i$). Both scores range from -1.0 (extremely polarized/not engaged at all) to +1.0 (not polarized at all/extremely engaged). Let $C$ be the PrimaryContent to which all agents are exposed. It consists of a friend list, trending, and imposed topics, each impacting polarization and engagement of $i$-th AgentPrompt.

This study assumes an initial state of polarization and engagement for all AgentPrompts, which can change during the simulation based on the following three functions:

- Plurality Function ($F_p$) is preferred when exposure to a wide array of diverging content is desired;
- Balanced Function ($F_b$) is selected when a balanced blend of similar and differing content is desired;
- Similarity Function ($F_s$) is preferred when exposure to homogenized content is desired.

Assuming polarization and engagement score functions for $i$-th AgentPrompt ($P_i$) at a state time $t$, i.e., $Ps_i(t)$, $Es_i(t)$, have parameters $\alpha, \beta \in [0.0, 1.0]$, respectively. The functions $F_p$, $F_b$, and $F_s$ represent the respective recommender system setups' impact on each AgentPrompt's polarization and engagement scores. Let $T_i$ denote the time factor for the daily activity of the $i$-th AgentPrompt. We can formulate polarization and engagement scores for agent $P_i$ at the state time $t + 1$ by the following equations:

$$Ps_i(t+1) = \alpha \cdot Ps_i(t) + (1-\alpha) \cdot F_p|F_b|F_s(P_i, C), \quad (2)$$

$$Es_i(t+1) = \beta \cdot Es_i(t) + (1-\beta) \cdot T_i \cdot F_p|F_b|F_s(P_i, C). \quad (3)$$

The functions' outcome depends on factors such as $P_i$'s initial attribute values, $C$'s complex nature, the interaction between $P_i$ and $C$, the duration of daily activity, and the relevant recommender system setup.

After simulating a set amount of state times (for instance, 365 days to simulate a yearly process), the final polarization and engagement score for each AgentPrompt $P_i$ (i.e., $Ps_i(365)$, $Es_i(365)$) can be used to measure the overall changes of polarization and engagement within the artificial social media network under each recommender system setup.

Through this mathematical simulation, we can measure how AgentPrompts' polarization and engagement levels change over time, providing a foundation for the effects of different recommender system setups on polarization and engagement in social media networks.

## 5. Experimental simulation

The experimental part of our study commenced with creating content intended for distribution within a simulated social network. This process involved generating content around three major issue-based themes (i.e., $C$'s imposed topics): Immigration Policies and Border Control, Climate Change and Environmental Regulations, and Healthcare System Reform. We produced five news articles for each issue through prompting, resulting in 15 news pieces. Each news article led to 10 related posts, culminating in a comprehensive set of 150 posts for the simulation [28].

Next, we established 22 simulated social network profiles, which we referred to as AgentPrompts [29]. We crafted these profiles based on 13 distinct descriptive, static, and dynamic dimensions described in Table 1.

The subsequent phase involved the selection of posts to which the AgentPrompts would be exposed [30]. We focused on three specific AgentPrompts for preliminary analysis. To achieve this, we employed a structured prompting approach that included detailed descriptions of the AgentPrompt profiles, a comprehensive list of the 150 posts, and delineation into three scenario categories: Plurality, Balanced, and Similarity. This methodology allowed us to curate 30 individual post recommendations for each AgentPrompt, ensuring each received a unique mixture of posts. The aim was to observe the effects of varied post recommendations on engagement and polarization metrics.

In the final step, we executed the social network simulation, termed RecSysLLMsP. We created three separate prompts tailored to the selected AgentPrompts [31, 32, 33]. Our main objective was to measure the reactions of AgentPrompts, specifically coded as PROFILE 1, PROFILE 21, and PROFILE 22, when they were exposed to a total of 90 posts. These posts were divided into three scenarios: Plurality, Balanced, and Similarity, with each scenario comprising 30 posts.

As indicated, we selected three (from 22) distinct AgentPrompts for this preliminary research to understand their reactions to different content mixes:

- PROFILE 1 was represented by *ArtLoverAnna*, an aspiring artist with a love for painting and museum visits [31]. Her personality profile indicated high levels of openness and agreeableness, paired with moderate conscientiousness and extraversion. Her dynamic dimensions recorded a slightly liberal political attitude, a high level of social connectivity, and a strong engagement level. Emotively, she was moderately reactive.
- PROFILE 21, *TruthSeekerTom*, was characterized by his nationalist stance and interest in conspiracy theories [32]. He exhibited lower openness and agreeableness but high neuroticism and a strong cognitive style leaning towards emotional thinking. His political attitude was extremely conservative, and he displayed both high social connectivity and engagement levels, coupled with intense emotional reactivity.
- Lastly, PROFILE 22 was *ActivistAlex*, an extremely liberal agent passionate about social justice and progressive causes [33]. Alex's personality traits reflected the highest levels of openness, agreeableness, and extraversion. He had a very open-minded cognitive style and a substantially liberal political attitude and exhibited both high social connectivity and engagement levels with moderate emotional reactivity.

It is important to note that we did not disclose whether the mix of posts aligned with the profiles' dimensions (i.e., whether they were plural, balanced, or similar). Instead, we presented detailed descriptions of each AgentPrompt, explaining their dimensions and the context behind those dimensions. Subsequently, we asked the AgentPrompts to react to the posts by either doing nothing or choosing an available reaction option such as *Like*, *Love*, *Care*, *Haha*, *Wow*, *Angry*, or *Sad*. Additional interaction options included commenting, sharing, and sending a friend request to the post's author. To answer the research questions, we take the Total Reactions count as a measure of engagement. At the same time, we use the Total Negative count (the sum of *Sad* and *Angry* reactions) to measure polarization.

Finally, we requested each AgentPrompt to self-report any changes in their dynamic dimensions due to the specific mix of posts they encountered. This approach aimed to evaluate the influence of content recommendations on their engagement levels and any potential shifts in their political attitudes. Through this methodology, we sought insights into how different content mixtures affect user engagement and polarization within a simulated social network environment.

Detailed findings of this research can be found in the OSF Repository [34].

# 6. Results

The results of our study are presented based on three distinct recommendation scenarios: Plurality, Balanced, and Similarity. Each scenario aimed to assess how varying content recommendations affected the engagement and reactions of three selected AgentPrompts: PROFILE 1 (*ArtLoverAnna*), PROFILE 21 (*TruthSeekerTom*), and PROFILE 22 (*ActivistAlex*).

In the Plurality Scenario (Table 2), where content recommendations were diverse and not necessarily aligned with the preferences of the AgentPrompts, the overall engagement patterns varied significantly across profiles. PROFILE 1 (*ArtLoverAnna*) recorded 48 interactions, with a markedly positive tilt (23 positive reactions compared to 1 negative). The *Love* reaction was predominant, suggesting strong affinity and engagement with varied content. PROFILE 21 (*TruthSeekerTom*) showed 16 interactions evenly split between positive and negative reactions (4 positive and 4 negative). This profile demonstrated a notable tendency for polarized reactions, particularly *Angry* responses. PROFILE 22 (*ActivistAlex*) displayed the highest engagement with 54 total interactions, split between high emotive and empathetic engagements (13 positive and 14 negative). *Care* and *Love* were the most frequent reactions, indicating a robust engagement with diverse content. Cumulatively, the Plurality Scenario resulted in 118 total reactions, with a noticeable inclination toward positive interactions (40) over negative ones (19).

A moderate but significant engagement pattern emerged in the Balanced Scenario (Table 3), where content recommendations blended similar and dissimilar content concerning the AgentPrompts' preferences. PROFILE 1 exhibited 42 total reactions, maintaining a positive engagement trend (20 positive reactions against 1 negative), with *Love* continuing to be the dominant reaction. PROFILE 21 provided 34 reactions, showing a mix of positive and negative feedback (12 positive vs. 5 negative). Notably, PROFILE 21 also engaged significantly through comments and *Angry* reactions. PROFILE 22 showed heightened activity with 57 total reactions, emphasizing emotional and empathetic engagement (26 positive and 3 negative), predominantly through *Care* and *Love* reactions. Overall, the Balanced Scenario resulted in 133 total reactions, with a decisive increase in positive interactions (58) compared to negative interactions (9).

In the Similarity Scenario (Table 4), where content closely aligned with the AgentPrompts' preferences, the engagement levels peaked significantly, reflecting strong alignment and affinity. PROFILE 1 recorded 50 total reactions, exclusively positive (25), illustrating high engagement through *Love* and *Care*. PROFILE 21 showed 30 total reactions, with a balanced distribution of responses (10 positive vs. 5 neg-ative), indicating engagement through preferred and resonant content. PROFILE 22 recorded the highest engagement with 59 total reactions, predominantly positive (29), indicating strong emotive connectivity through *Love* and *Care* reactions, and minimum negative reactions. The Similarity Scenario led to a total of 139 interactions, weighted heavily towards positive responses (64) versus negative ones (6).

**Table 2**
Plurality Scenario: recommending different content when compared to preferences of AgentPrompts

|  | Haha | Like | Wow | Care | Love | Sad | Angry | Comments | Total Reactions | Total Positive | Total Negative |
|---|---|---|---|---|---|---|---|---|---|---|---|
| PROFILE 1 | 0 | 0 | 4 | 3 | 16 | 1 | 0 | 24 | 48 | 23 | 1 |
| PROFILE 21 | 0 | 4 | 0 | 0 | 0 | 0 | 4 | 8 | 16 | 4 | 4 |
| PROFILE 22 | 0 | 0 | 1 | 8 | 4 | 0 | 14 | 27 | 54 | 13 | 14 |
| In total |  |  |  |  |  |  |  |  | 118 | 40 | 19 |

**Table 3**
Balanced Scenario: recommending both similar and different content when compared to preferences of AgentPrompts

|  | Haha | Like | Wow | Care | Love | Sad | Angry | Comments | Total Reactions | Total Positive | Total Negative |
|---|---|---|---|---|---|---|---|---|---|---|---|
| PROFILE 1 | 0 | 0 | 2 | 3 | 15 | 1 | 0 | 21 | 42 | 20 | 1 |
| PROFILE 21 | 0 | 4 | 0 | 0 | 8 | 0 | 5 | 17 | 34 | 12 | 5 |
| PROFILE 22 | 0 | 0 | 1 | 7 | 18 | 0 | 3 | 28 | 57 | 26 | 3 |
| In total |  |  |  |  |  |  |  |  | 133 | 58 | 9 |

The engagement data across all scenarios illustrate patterns reflective of each AgentPrompt's personality and dynamic dimensions. The Plurality Scenario elicited diverse responses with a leaning towards positive engagement but varied significantly across profiles. The Balanced Scenario struck a moderate middle ground, increasing overall positive engagement while maintaining low negative reactions. The Similarity Scenario maximized engagement and minimized negative responses, showing the highest alignment with user preferences.

These results suggest that content recommendations tailored to reflect user preferences (Similarity Scenario) foster higher engagement and positive interaction. In contrast, diverse content (Plurality Scenario) stimulates mixed reactions with a balanced trajectory of engagement in the Balanced Scenario. This insight is critical in understanding the dynamics of social network interactions and the impact of content recommendations on user engagement and polarization.

## 7. Conclusions

The development of RecSysLLMsP introduces a systematic way to assess the societal implications of recommender systems on social polarization and engagement. As social media increasingly becomes the primary source of news, information, and interaction, studying the potential effects of recommender systems becomes paramount both for the overall democratic well-being of societies and the business models of social media companies.

Our study explores the impacts of different recommender system setups on community engagement and polarization within a simulated social network. Through our novel simulation software, RecSysLLMsP, we developed and utilized diverse AI agents called AgentPrompts to simulate the user experience. The three scenarios we tested—Plurality, Balanced, and Similarity—provided rich insights into the dynamic interplay between content recommendations and user engagement.

The findings indicate that variations in recommender system setups significantly influence community engagement and polarization. The Similarity Scenario, which aligned content closely with user preferences, resulted in the highest levels of engagement and predominantly positive interactions. This result suggests that content tailored to individual preferences fosters a more engaging and less divisive community environment. In contrast, the Plurality Scenario, which recommended varied content, elicited a more mixed response, highlighting higher positive and negative engagement levels. The Balanced Scenario achieved a moderate level of engagement, showing that a mixture of similar and varied content can promote a balanced user experience with reduced polarization.

The mathematical model of the simulation was outlined using these scenarios, permitting a systematic examination of user engagement and polarization changes over simulated time. Initial results indicate the significant influence of the recommender systems on users' engagement and polarization.

These results have broader societal implications. Recommender systems favouring highly personalized content may enhance user satisfaction and engagement but could also contribute to echo chambers and social bubbles, potentially exacerbating societal polarization. Conversely, systems that offer a plurality of perspectives might mitigate such polarization but at the cost of reduced user satisfaction and engagement. Our findings underscore the need for a careful balance in designing recommender systems to align user preferences with the societal goal of reducing polarization. Nevertheless, the results also demonstrate the ability of generative AI to serve as a content curator tailored to individual user preferences, potentially increasing user engagement and polarization, thus boosting both the positive and negative effects of recommender systems.

While our study provides valuable insights, it is not without limitations. The simulated environment may not fully capture the complexities of real-world social networks. The AgentPrompts, while diverse, are simplifications of real human users and do not encompass the full range of human behaviour and unpredictability. Additionally, the content generated for our simulation was based on specific issues, which might not reflect the broad spectrum of topics encountered in actual social networks. Hence, extrapolating

**Table 4**
Similarity Scenario: recommending similar content when compared to preferences of AgentPrompts

|  | Haha | Like | Wow | Care | Love | Sad | Angry | Comments | Total Reactions | Total Positive | Total Negative |
|---|---|---|---|---|---|---|---|---|---|---|---|
| PROFILE 1 | 0 | 0 | 0 | 3 | 22 | 0 | 0 | 25 | 50 | 25 | 0 |
| PROFILE 21 | 0 | 2 | 0 | 0 | 8 | 0 | 5 | 15 | 30 | 10 | 5 |
| PROFILE 22 | 0 | 0 | 3 | 6 | 20 | 0 | 1 | 29 | 59 | 29 | 1 |
| In total |  |  |  |  |  |  |  |  | 139 | 64 | 6 |

our findings to real-world scenarios should be cautiously approached.

Future research should address these limitations by incorporating more complex and realistic user models and expanding the range of content topics. Longitudinal studies could provide deeper insights into the long-term effects of different recommender system setups on user engagement and societal polarization. Incorporating real user data while ensuring privacy and ethical considerations could enhance the validity and applicability of the findings. Exploring the role of user education and awareness in mitigating the negative effects of both echo chambers and information overload presents another valuable research direction. Ultimately, advancing our understanding of the optimal balance between personalized content and diverse perspectives will contribute to developing more responsible and socially beneficial recommender systems.

RecSysLLMsP posits a significant advance in recommender systems, providing a robust platform to thoroughly evaluate the broader societal and engagement effects of different recommender system setups, helping make more informed decisions when implementing those systems. While this study provides a pioneering method to simulate social network dynamics, it also highlights the need for future research entailing long-term simulations. Such studies would offer a deeper understanding of how prolonged exposure to specific recommender system setups impacts user engagement and polarization.

## Acknowledgments

This paper has been supported by the TWON (project number 101095095), a research project funded by the European Union under the Horizon Europe framework (HORIZON-CL2-2022-DEMOCRACY-01, topic 07). More details about the project can be found on its official website: https://www.twon-project.eu/.

This research has been accomplished with the support and collaboration of the COST Action Network CA21129 - What are Opinions? Integrating Theory and Methods for Automatically Analyzing Opinionated Communication (OPINION) - https://www.opinion-network.eu/.

This paper was realized with the support of the Ministry of Science, Technological Development and Innovation of the Republic of Serbia, according to the Agreement on the realization and financing of scientific research 451-03-66/2024-03/200025.

## References

[1] H. Allcott, M. Gentzkow, Social media and fake news in the 2016 election, Journal of economic perspectives 31 (2017) 211–236.

[2] C. Sunstein, # Republic: Divided democracy in the age of social media, Princeton university press, 2018.

[3] S. Flaxman, S. Goel, J. M. Rao, Filter bubbles, echo chambers, and online news consumption, Public opinion quarterly 80 (2016) 298–320.

[4] T. Donkers, J. Ziegler, The dual echo chamber: Modeling social media polarization for interventional recommending, in: Proceedings of the 15th ACM conference on recommender systems, 2021, pp. 12–22.

[5] P. Ramaciotti Morales, J.-P. Cointet, Auditing the effect of social network recommendations on polarization in geometrical ideological spaces, in: Proceedings of the 15th ACM Conference on Recommender Systems, 2021, pp. 627–632.

[6] K. Shivaram, P. Liu, M. Shapiro, M. Bilgic, A. Culotta, Reducing cross-topic political homogenization in content-based news recommendation, in: Proceedings of the 16th ACM Conference on Recommender Systems, 2022, pp. 220–228.

[7] W. G. Mangold, D. J. Faulds, Social media: The new hybrid element of the promotion mix, Business horizons 52 (2009) 357–365.

[8] J. Van Dijck, T. Poell, Understanding social media logic, Media and communication 1 (2013) 2–14.

[9] E. Pariser, The filter bubble: What the Internet is hiding from you, penguin UK, 2011.

[10] Y. Deldjoo, D. Jannach, A. Bellogin, A. Difonzo, D. Zanzonelli, Fairness in recommender systems: research landscape and future directions, User Modeling and User-Adapted Interaction 34 (2024) 59–108.

[11] Y. Deldjoo, Z. He, J. McAuley, A. Korikov, S. Sanner, A. Ramisa, R. Vidal, M. Sathiamoorthy, A. Kasrizadeh, S. Milano, et al., Recommendation with generative models, arXiv preprint arXiv:2409.15173 (2024).

[12] Y. Deldjoo, Z. He, J. McAuley, A. Korikov, S. Sanner, A. Ramisa, R. Vidal, M. Sathiamoorthy, A. Kasirzadeh, S. Milano, A review of modern recommender systems using generative models (Gen-RecSys), in: Proceedings of the 30th ACM SIGKDD Conference on Knowledge Discovery and Data Mining, 2024, pp. 6448–6458.

[13] A. Bruns, After the 'apicalypse': Social media platforms and their fight against critical scholarly research., Information, Communication & Society (2019).

[14] J. A. Tucker, Y. Theocharis, M. E. Roberts, P. Barberá, From liberation to turmoil: Social media and democracy, J. Democracy 28 (2017) 46.

[15] C. A. Bail, L. P. Argyle, T. W. Brown, J. P. Bumpus, H. Chen, M. F. Hunzaker, J. Lee, M. Mann, F. Merhout, A. Volfovsky, Exposure to opposing views on social media can increase political polarization, Proceedings of the National Academy of Sciences 115 (2018) 9216–9221.

[16] L. Bojic, Metaverse through the prism of power and

addiction: what will happen when the virtual world becomes more attractive than reality?, European Journal of Futures Research 10 (2022) 22.

[17] L. Bojic, Ai alignment: Assessing the global impact of recommender systems, Futures (2024) 103383.

[18] S. Alipour, A. Galeazzi, E. Sangiorgio, M. Avalle, L. Bojic, M. Cinelli, W. Quattrociocchi, Cross-platform social dynamics: an analysis of chatgpt and covid-19 vaccine conversations, Scientific Reports 14 (2024) 2789.

[19] N. J. Foss, T. Saebi, Fifteen years of research on business model innovation: How far have we come, and where should we go?, Journal of management 43 (2017) 200–227.

[20] Y. A. Kim, J. Srivastava, Impact of social influence in e-commerce decision making, in: Proceedings of the ninth international conference on Electronic commerce, 2007, pp. 293–302.

[21] K.-C. Yang, O. Varol, C. A. Davis, E. Ferrara, A. Flammini, F. Menczer, Arming the public with artificial intelligence to counter social bots, Human Behavior and Emerging Technologies 1 (2019) 48–61.

[22] T. B. Brown, B. Mann, N. Ryder, M. Subbiah, J. Kaplan, P. Dhariwal, A. Neelakantan, P. Shyam, G. Sastry, A. Askell, S. Agarwal, A. Herbert-Voss, G. Krueger, T. Henighan, R. Child, A. Ramesh, D. M. Ziegler, J. Wu, C. Winter, C. Hesse, M. Chen, E. Sigler, M. Litwin, S. Gray, B. Chess, J. Clark, C. Berner, S. McCandlish, A. Radford, I. Sutskever, D. Amodei, Language models are few-shot learners, in: Proceedings of the 34th International Conference on Neural Information Processing Systems, NIPS '20, Curran Associates Inc., Red Hook, NY, USA, 2020.

[23] E. M. Bender, T. Gebru, A. McMillan-Major, S. Shmitchell, On the dangers of stochastic parrots: Can language models be too big?, in: Proceedings of the 2021 ACM Conference on Fairness, Accountability, and Transparency, FAccT '21, Association for Computing Machinery, New York, NY, USA, 2021, p. 610–623. URL: https://doi.org/10.1145/3442188.3445922. doi:10.1145/3442188.3445922.

[24] L. Lin, L. Wang, J. Guo, K.-F. Wong, Investigating bias in llm-based bias detection: Disparities between llms and human perception, arXiv preprint arXiv:2403.14896 (2024).

[25] A. Taubenfeld, Y. Dover, R. Reichart, A. Goldstein, Systematic biases in llm simulations of debates, arXiv preprint arXiv:2402.04049 (2024).

[26] H. Zhou, Z. Feng, Z. Zhu, J. Qian, K. Mao, Unibias: Unveiling and mitigating llm bias through internal attention and ffn manipulation, arXiv preprint arXiv:2405.20612 (2024).

[27] L. Bojić, M. Cinelli, D. Ćulibrk, B. Delibašić, Cern for ai: a theoretical framework for autonomous simulation-based artificial intelligence testing and alignment, European Journal of Futures Research 12 (2024) 1–19.

[28] OpenAI, RecSysLLMsP - 3 Issues, 15 news (5 per issue), 150 posts (10 per news), https://platform.openai.com/playground/p/skjaoRqUqoayVWRW65U2zqPE?model=undefined&mode=chat, 2024. [Online; accessed 29-August-2024].

[29] OpenAI, RecSysLLMsP-AgentPrompts (20), https://platform.openai.com/playground/p/PMmYZuxIMJxsRt3ZGO5tXCZJ?model=undefined&mode=chat, 2024. [Online; accessed 29-August-2024].

[30] OpenAI, RecSysLLMsP-Selection of social media posts for each of 3 scenarios for 3 agents for initial experiment, https://platform.openai.com/playground/p/KA0WQbkcUEpzPtRCnKnPIF4D?model=undefined&mode=chat, 2024. [Online; accessed 29-August-2024].

[31] OpenAI, RecSysLLMsP-Measuring reactions of one AgentPrompt PROFILE 1 exposed to 90 posts in 3 scenarios, https://platform.openai.com/playground/p/vsQDwa3UWDt50kXme4DqB1Fi?model=undefined&mode=chat, 2024. [Online; accessed 29-August-2024].

[32] OpenAI, RecSysLLMsP-Measuring reactions of one AgentPrompt PROFILE 21 exposed to 90 posts in 3 scenarios, https://platform.openai.com/playground/p/lgkIomAbiZy95Z7zTYgKLuoR?model=undefined&mode=chat, 2024. [Online; accessed 29-August-2024].

[33] OpenAI, RecSysLLMsP-Measuring reactions of one AgentPrompt PROFILE 22 exposed to 90 posts in 3 scenarios, https://platform.openai.com/playground/p/WM91bLWTtJY5pnpfno0ma7D8?model=undefined&mode=chat, 2024. [Online; accessed 29-August-2024].

[34] O. S. F. O. Repository, Towards Recommender Systems LLM Playground (RecSysLLMsP): Exploring Polarization and Engagement in Simulated Social Networks, https://osf.io/qsrmx/?view_only=14d3858cd4684e63ad7a928e934e8e9b, 2024. [Online; accessed 29-August-2024].

## A. Online Resources

The teaser video of this paper is available on the following link: RecSysLLMsP